\documentstyle[preprint,revtex]{aps}

\ifx\selectfont\undefined 

   \newcommand{\BM}[1]{{\bf #1}}
   \newcommand{\BMIT}[1]{{\bf #1}}
   \newcommand{\skript}[1]{{\cal #1}}

   \newcommand{\mathrm}[1]{{\rm #1}}
   \newcommand{\normalshape}{\rm }

\else 

   \newmathalphabet{\BM}
   \addtoversion{normal}{\BM}{cmr}{bx}{n}

   \newmathalphabet{\BMIT}
   \addtoversion{normal}{\BMIT}{cmm}{b}{it}

   \newmathalphabet{\skript}
   \addtoversion{normal}{\skript}{eus}{m}{n}

\fi

\newcommand{\be}{\begin{equation}}
\newcommand{\ee}{\end{equation}}

\newcommand{\bea}{\begin{eqnarray}}
\newcommand{\eea}{\end{eqnarray}}

\newcommand{\BA}[1]{\begin{array}{#1}\displaystyle }
\newcommand{\BAT}[1]{\begin{array}[t]{#1}\displaystyle }
\newcommand{\BAB}[1]{\begin{array}[b]{#1}\displaystyle }
\newcommand{\CR}{\\ \noalign{\vskip2pt} \displaystyle }
\newcommand{\XX}{& \displaystyle }
\newcommand{\EA}{\end{array}}

\newcommand{\Vec}[1]{\BM{#1}}

\newcommand{\middel}[1]{\left\langle #1 \right\rangle }
\newcommand{\braket}[1]{\left\langle #1 \right\rangle }
\newcommand{\half}{{1 \over 2}}
\def\quarter{{1 \over 4}}
\newcommand{\spinone}{\mbox{spin--$1$} }

\newcommand{\susc}{\mbox{\Large $\chi$}}
\newcommand{\kb}{k_{\mathrm{B}}}
\newcommand{\kbT}{\kb T }

\begin{document}
\draft

\begin{title}
Phase Diagram of a Lattice Microemulsion Model \\ in Two Dimensions
\end{title}

\author{Per Arne Slotte}
\begin{instit}
           Gruppe for teoretisk fysikk \\
           Institutt for fysikk --- NTH \\
           N--7034 Trondheim \\
           Norway
\end{instit}

\begin{abstract}
   The phase diagram of a lattice microemulsion model
   proposed by Ciach, H{\o}ye and Stell\cite{CHS-1,CHS-2}
   is studied using mean-field theory and Monte Carlo simulations.
   Surfactant directional degrees of freedom are summed out exactly
   before mean-field theory is applied, and the resulting
   phase diagrams are much improved compared with previous
   mean-field results. The critical line and tricritical
   point is located using Monte Carlo simulations and finite size
   scaling.
\end{abstract}

\section{Introduction}

Oil, water, and surfactant mixtures exhibit very special
properties that makes them interesting from a
practical as well as from a theoretical point of view.
Under certain circumstances they form
microemulsions --- phases in which microscopic
oil- and water-occupied regions are separated by thin layers
of surfactant. The characteristic length scale of
this structure is typically much larger than the
``natural'' length scale determined from the particle interactions,
while still microscopic.

This paper discusses the phase diagram of the two-dimensional
version of a lattice model for a
three-component microemulsion system proposed by
Ciach, H{\o}ye and Stell\cite{CHS-1,CHS-2}, and studied in
subsequent papers\cite{CHS-3,CHS-4,CHS-5,CHS-6,CHS-7,Laradji}. In the following
I will refer to this model as the CHS model. Figure~\ref{fig1}(a)--(c) shows
the three types of phase diagram topologies that is
expected\cite{CHS-4,CHS-7,Laradji}
for different surfactant strengths. In the following I will
refer to these phase diagram topologies as type A, B and C respectively.
The phase diagrams are
given in the temperature/surfactant-chemical-potential plane for
equal oil/water chemical potentials. With a weak surfactant,
the phase diagram is of the simple three-component type shown in
figure~\ref{fig1}(a), with two
regions, disordered and oil/water coexistence, separated by
a line of continuous transitions which changes into a first-order line
at a tricritical point. When the surfactant is strong, there are
two main possibilities shown in figure~\ref{fig1}(b) and (c).
Both diagrams have four regions:
disordered, microemulsion, oil/water coexistence, and
ordered structured phases (incommensurate, layered, bicontinuous \ldots).
The disordered- and microemulsion regions are separated by a
Lifshitz line\cite{Hornreich,Schick-4}
where the peak in the water--water structure
function moves away from zero wave vector, and not by a phase
transition.
Note that some workers use the term disorder line for the
Lifshitz line. I use here the nomenclature of Gompper and
Schick\cite{Schick-4}, which discriminates between the
Lifshitz line, and the disorder line  at which the
asymptotic decay of the water--water correlation function
changes from monotonic to nonmonotonic.
In phase diagrams of type B,
the disordered region is separated from the oil/water
coexistence region by a line of continuous phase transitions. The
transition
changes into first order at a tricritical point. The Lifshitz line
intersects the first-order line, and thus we have
the characteristic oil/water/microemulsion coexistence
observed experimentally in many systems. The microemulsion
is separated from the ordered structured phases by lines of
phase transitions that may be either continuous or first order.
In phase diagrams of type C,
the line separating disordered from oil/water
coexistence is always a line of continuous phase transitions, and the
line ends at a
Lifshitz point where it intersects the Lifshitz line. Thus
there is no oil/water/microemulsion coexistence.
The microemulsion is separated from an ordered
structured phase by a line of continuous transitions near the Lifshitz
point, but the transition may be of any order elsewhere.
The range of parameters where phase diagrams of type C is found
is substantially reduced in the present calculations, compared
to earlier mean-field calculations\cite{CHS-4,CHS-7}.

   The article is organized as follows:
In section~\ref{sec:model} the CHS hamiltonian is rewritten in
terms of several \spinone Ising variables on each site.
In section~\ref{sec:spinone} an exact mapping of the model onto an
ordinary \spinone model
with temperature-dependent (effective) multispin couplings is developed.
The connection  between the CHS model and the
\spinone model of Schick and Shih\cite{Schick-2} is also discussed.
In sections~\ref{sec:MF} and~\ref{sec:correl}
standard mean-field theory is applied to the
effective \spinone model.
Results from Monte Carlo simulations are presented in section~\ref{sec:MC}.
In section~\ref{sec:conc} the resulting
phase diagrams are discussed and compared with phase diagrams
obtained in related works.


\section{The model}
\label{sec:model}

Each site in a hypercubic $d$-dimensional lattice is occupied by oil, water
or surfactant. The surfactant can take $2 d$ different orientations
along the lattice directions. This gives a total of $2+2 d$ states
for each site.

The hamiltonian may be written in terms of $1+d$ \spinone Ising variables
per site:
$\sigma$ and $\tau^{x_i}$, where $x_i$ is a lattice direction.
$\sigma = \pm 1$ represents oil/water, $\sigma = 0$ surfactant,
and the $\tau$'s different surfactant directions. Note that
these variables are {\em dependent\/} since one, and only one, of the variables
at a given site will take a nonzero value. This gives again a total of $2+2 d$
states.

For simplicity it is assumed
that the hamiltonian is symmetric with respect to interchanging
oil with water and flipping the surfactant ends, ie $\sigma \to -\sigma$ and
$\tau \to -\tau$. To be able to regulate the oil/water concentration ratio,
the hamiltonian must in addition include an unsymmetric field term. If we
include nearest-neighbor interactions only, and
exclude direction dependent surfactant--surfactant interactions, the most
general form of the hamiltonian is:
\be
 \skript{H} = - \sum_i \left( H\sigma_i + \mu\sigma_i^2 \right)
            - \sum_{\braket{ij}} \left( J \sigma_i\sigma_j
                                      + K \sigma_i^2\sigma_j^2
                                      + A \left(
\sigma_i\tau_j^{x_{ij}^{\parallel}}
                                          - \tau_i^{x_{ij}^{\parallel}}\sigma_j
\right)
                                 \right) \; .
  \label{eq:model-hamiltonian}
\ee
Here the first sum runs over all sites, while the second sum runs over all
pairs
$\braket{ij}$ of nearest-neighbor sites, and $x_{ij}^{\parallel}$ is the
lattice
direction parallel with the bond $\braket{ij}$. The physical significance
of the parameters in the hamiltonian are: $H$ regulates the oil/water
concentration ratio. $\mu$ regulates the surfactant density. $J>0$
ensures that oil and water do not mix at low temperatures, and
will be used as a temperature scale. $K$ is the isotropic
surfactant--surfactant interaction, I will assume $K\ge 0$. $A$ is the
surfactant strength, the sign of $A$ is physically irrelevant. Note
the antisymmetric form of the $A$-term, this reflects the
amphiphilic nature of the surfactant (ie.\ that one end loves water
and the other end loves oil). This term is the only term
in the hamiltonian that is new compared with a simple hamiltonian for
a three-component system with isotropic interactions.

The hamiltonian has in other
work\cite{CHS-1,CHS-2,CHS-3,CHS-4,CHS-5,CHS-6,CHS-7}
been formulated in
terms of the oil--oil (water--water) energy,
$-b$, the oil--surfactant energy, $\pm c$, and the chemical
potentials for oil, $\mu_1$, water, $\mu_2$,  and surfactant,
$\mu_3=\mu_4=\mu_5=\mu_6=\mu_s$. The correspondence between
the parameters in the hamiltonian (\ref{eq:model-hamiltonian}) above
and the parameters $b$, $c$, $\mu_1$,
$\mu_{2}$, and $\mu_s$, is as follows:
\bea
    H &=& \half (\mu_{1} - \mu_{2}) \nonumber \\
    J &=& \half b \nonumber \\
    A &=& c \\
 \label{eq:model-param-CHS}
    \mu &=& \mu_s - \half (\mu_{1} + \mu_{2}) \nonumber \\
    K &=& \half b \nonumber
\eea

The ground states, and $T=0$ phase diagram, of the model
have been previously analyzed for the $K=J$ case\cite{CHS-2},
and for all $K>-J$ the structure is the same. This is
summarized in figure~\ref{fig1.0.1}.

The present spin-language formulation may not be appropriate if direction
dependent surfactant--surfactant interactions are important, since the
inclusion of such terms will complicate the hamiltonian
(\ref{eq:model-hamiltonian})
substantially.

It should be noted that
the purpose of the present lattice model is not to accurately model the
phase diagram of the real oil--water--surfactant system.
A simple model of this type does not even model the
{\em two component\/} (oil--water) system correctly:
Due to the complexity of the oil--oil, water--water and
oil--water interactions a temperature-dependent coupling
$J$ is needed to reproduce the important property
called reentrant solubility,\cite{Schick-3,Goldstein} i.e.\ that at fixed
concentration oil an water mixes at high temperatures, and phase
separates below a certain temperature, but mixes again
in a range of lower temperatures. This stems from the complex nature of
the bonds which can be in one of several van der Waals and hydrogen
bonding states. Moreover, oil and
water are certainly not symmetric in their surfactant interactions.
The aim is rather to study the simplest possible model to be able to
isolate the basic ingredients that are necessary to reproduce the key
properties of real microemulsions.


\section{Correspondence with spin-$1$ Ising model}
\label{sec:spinone}

The CHS model is closely related to two other lattice models
proposed for microemulsion systems. One of these is the
model proposed by
Matsen and Sullivan\cite{Matsen-1,Matsen-2} which is a
generalization of the CHS model. The other is
the \spinone Ising model proposed by Schick and
Shih\cite{Schick-2} and studied in a series of subsequent
papers \cite{Schick-4,Schick-3,Schick-1,Schick-6,Schick-5,Schick-7}. In the
following I will call this model the Schick model.

It is clear from
the form of the hamiltonian (\ref{eq:model-hamiltonian})
that the Schick model and the CHS model  are closely related
since both are extensions of the
basic \spinone Blume-Emery-Griffith (BEG) model:
\be
 \skript{H}_{\mathrm{BEG}} =
            - \sum_i \left( H\sigma_i + \mu\sigma_i^2 \right)
            - \sum_{\braket{ij}} \left( J \sigma_i\sigma_j
                                      + K \sigma_i^2\sigma_j^2
                                 \right) \; .
 \label{eq:spinone-BEGhamiltonian}
\ee
In the Schick model, the microemulsion character  arises from an
additional three particle interaction
$L\sigma_i (1-\sigma_j^2) \sigma_k$, where the sites $ijk$ are
neighbors on a line. The CHS model is more ``realistic'' with
directed surfactants and pair interactions. As long as no direction
dependent surfactant--surfactant interactions are included in
the CHS model, the directional degrees of freedom can be summed
out independently on each site resulting in an effective \spinone
hamiltonian of the Schick type with temperature-dependent multiple
site couplings\cite{Schick-3,Alexander-1,Halley}.
Hence, there exist an exact mapping between the
CHS model and an extended version of the Schick model. Below, this
mapping is
explicitly calculated for the one dimensional version
of the CHS model. The explicit expression for the effective
hamiltonian in two dimensions is given in appendix A.

The CHS hamiltonian (\ref{eq:model-hamiltonian}) has the following symmetries
that also must be valid for the effective \spinone  hamiltonian:
\begin{itemize}
\item  A global spin flip $\sigma \to -\sigma, \; H \to -H \; \tau \to -\tau$,
       which implies
       that all terms must be even in $\sigma$ operators.
\item  Point group operations (In one dimension limited to
       reflections around a point).
\end{itemize}
Summing out the surfactant directions on site $i$ gives rise to
terms proportional to $(1-\sigma_i^2)$ and involving the $\sigma$'s
on the
neighboring sites. In one dimension, four terms of this type are
consistent with the above symmetries:
\be
  \BA{l}
    X_0 (1-\sigma_i^2) \CR
    X_1 (1-\sigma_i^2)(\sigma_{i-1}^2 + \sigma_{i+1}^2) \CR
    X_2 (1-\sigma_i^2)\sigma_{i-1}\sigma_{i+1} \CR
    X_3 (1-\sigma_i^2)\sigma_{i-1}^2\sigma_{i+1}^2
  \EA \; .
  \label{eq:spinone-1D-terms}
\ee
$X_2$ corresponds to $L$ in the original Schick hamiltonian.
Considering triplets of sites, the following four equations emerge from summing
out
the directional degree of freedom at the central cite,
\be
  \exp(-\beta\skript{H}_{\mathrm{eff}}) =
    \sum_\tau \exp(-\beta\skript{H}) \; ,
  \label{eq:spinone-summing}
\ee
with $\beta = 1/\kbT$, $T$ temperature,
and $\kb$ the Boltzmann constant:
\be
  \BA{lrcl}
    \oplus - \odot - \oplus  : \,\XX  X_2 + X_3 + 2 X_1 + X_0 \XX =\XX
          -\kbT\ln 2 \CR
    \ominus - \odot - \oplus : \,\XX -X_2 + X_3 + 2 X_1 + X_0 \XX =\XX
          -\kbT\ln(e^{2\beta A}+e^{-2\beta A}) \CR
    \odot - \odot - \oplus   : \,\XX              X_1 + X_0 \XX =\XX
          -\kbT\ln(e^{\beta A}+e^{-\beta A}) \CR
    \odot - \odot - \odot    : \,\XX                    X_0 \XX =\XX
          -\kbT\ln 2
  \EA\; .
  \label{eq:spinone-1D-equations}
\ee
Here the symbols $\ominus$, $\odot$, and $\oplus$ represent $\sigma = -1$,
$\sigma = 0$, and $\sigma = +1$, respectively. Solving the linear equations
(\ref{eq:spinone-1D-equations}) is trivial:
\bea
    X_0 &=& - \kbT \ln 2 \nonumber\\*
    X_1 &=& -A - \kbT \ln(1+e^{-2\beta A})
 \label{eq:spinone-1D-solutions}\\*
    X_2 &=&  A + \half\kbT\ln(1+e^{-4\beta A}) \nonumber\\*
    X_3 &=&  A - \half\kbT\ln(\frac{1+e^{-4\beta A}}{(1+e^{-2\beta A})^4})
\nonumber
\eea
Note that the term proportional to $X_1$ in (\ref{eq:spinone-1D-terms}) splits
into
a single-site term and a pair term, and the term proportional
to $X_0$ splits into a constant and a single-site term.
Thus, in one dimension the effective \spinone hamiltonian is:
\be
  \BA{rl}
   \skript{H}_{\mathrm{eff}} = \sum_i \, ( \XX \!\!\!
                                   -H\sigma_i - (\mu+X_0-2 X_1)\sigma_i^2
                                   -J \sigma_i\sigma_{i+1}
                                   -(K+2X_1)\sigma_i^2\sigma_{i+1}^2  \CR
                              \XX \!\!\!   + (1-\sigma_i^2)(
               X_2 \sigma_{i-1}\sigma_{i+1}
             + X_3 \sigma_{i-1}^2 \sigma_{i+1}^2 ) + X_0 )
  \EA \; .
\label{eq:spinone-1D-eff}
\ee
This hamiltonian has an extra three-spin interaction,
$X_3 \sigma_{i-1}^2 (1-\sigma_i^2) \sigma_{i+1}^2$, compared with the original
Schick hamiltonian, and in higher dimensions even more interactions must
be included.

In the above expressions for the effective couplings
(equations~\ref{eq:spinone-1D-solutions} and~\ref{eq:spinone-2D-solutions}),
the first term is the zero-tem\-pe\-ra\-ture value, and the temperature
dependence is isolated in the second term.
With exception of the entropic single-site $X_0$ coupling,
all the effective couplings have a simple monotonous temperature
dependence, and tend to zero with increasing temperature.
Since the temperature behavior of the
effective couplings in the $\mbox{CHS} \to \mbox{Schick}$ mapping
is rather trivial, many qualitative features of the CHS model
are probably preserved when one considers a \spinone
model with temperature independent couplings. Note, however,
the nontrivial behavior of the nearest neighbor
$\sigma_i^2\sigma_j^2$ coupling, $(K+2X_1)$, in the \spinone hamiltonian
(equation~\ref{eq:spinone-1D-eff}). For $A>K/2$ the coupling is
negative at low temperatures, and changes sign with increasing temperature.

The CHS model may be applied to the two component
oil--surfactant (or water--surf\-act\-ant) system ($H \to \pm\infty$),
and contains all the
necessary ingredients to get a
microemulsion like phase where water rich regions
are separated by surfactant bilayers\cite{CHS-6}.
In a study of this system using the Schick model\cite{Schick-3}, a four-site
interaction was added to get the bilayer structure.
The present calculation in one dimension
indicates that it may be possible to model the
bilayer structure with  a \spinone hamiltonian using
two- and three-site interactions only. The important ingredient is
probably the nontrivial $\sigma_i^2\sigma_j^2$ coupling
mentioned above.


\section{MF theory of the effective hamiltonian}
\label{sec:MF}

In this section standard mean-field theory is applied to the
two-dimensional version of the effective hamiltonian of
section \ref{sec:spinone}. In earlier
work\cite{CHS-3,CHS-4,CHS-5,CHS-7}
mean-field
theory was applied to the CHS hamiltonian (\ref{eq:model-hamiltonian})
directly, and I expect that the present approach will improve on
these results.

The starting point of mean-field theory is to express
the energy and entropy per spin as functionals
of the magnetization, $m = \middel{\sigma}$ ($\middel{\ldots}$
represents thermal average), and surfactant density,
$\rho_s = \middel{1-\sigma^2}$,
by assuming that the probabilities of the different states
are uncorrelated at different sites.
For {\em homogeneous phases\/} (all sites equivalent) the mean-field
energy per site,
$e$, is
\bea
   e & = &
  -H m   - 2 J {m^2} - \left( 1 - \rho_{s}  \right) \mu  -
   2 K {{\left( 1 - \rho_{s}  \right) }^2} + \nonumber\\
  & &
   \left( X_{0}
      + 4 X_{1} \left( 1 - \rho_{s}  \right)
      + 2 X_{2} {m^2}
      + 6 X_{3} {{\left( 1 - \rho_{s}  \right) }^2}
      + 4 X_{6} {m^2} \left( 1 - \rho_{s}  \right) + \right. \nonumber\\
  & &
   \phantom{(}\left. 4 X_{8} {{\left( 1 - \rho_{s}  \right) }^3}
      + X_{9} {m^4}
      + 2 X_{10} {m^2} {{\left( 1 - \rho_{s}  \right) }^2}
      + X_{12} {{\left( 1 - \rho_{s}  \right) }^4} \right)  \rho_{s}
  \; ,
  \label{eq:eff_MF_e}
\eea
where the effective couplings, $X_n$, are given by
equation~\ref{eq:spinone-2D-solutions}.
The entropy per site, $s$, takes the simple form
\bea
   s/\kb = - \frac{1 - m - \rho_{s}}{2}
         \ln (\frac{1 - m - \rho_{s}}{2} )  -
   \frac{1 + m - \rho_{s}}{2}  \ln ( \frac{1 + m - \rho_{s}}{2}) -
   \rho_{s}  \ln (\rho_{s} )
  \; .
  \label{eq:eff_MF_s}
\eea

Note that when mean-field theory is applied to the CHS hamiltonian
directly, the equivalents of equations \ref{eq:eff_MF_e} and
\ref{eq:eff_MF_s} are {\em independent} of the surfactant strength
$A$. Hence the
part of the phase diagram that pertains to homogeneous
phases, ie disordered to oil/water coexistence transitions
and tricritical point, is totally insensitive to the
surfactant strength in that approximation. The present
approach removes this artifact.

The free energy at a given field, $H$, and chemical potential,
$\mu$, $f(\beta,\mu,H)$, is found by minimizing
$f(\beta,\mu,H;\rho_s,m) = e - \beta^{-1} (s/\kb)$
with respect to $\rho_{s}$ and $m$.
Thus, the partial derivatives of $f$ with respect to $m$ and $\rho_{s}$
must vanish:
\bea
    \frac{\partial f}{\partial m} & = &
   -H - 4 J m + \left( 4 X_{2} m + 4 X_{9} {m^3} +
      8 X_{6} m \left( 1 - \rho_{s}  \right)  +
      4 X_{10} m {{\left( 1 - \rho_{s}  \right) }^2} \right)  \rho_{s}
  \nonumber\\
  & &
   -\frac{1}{2\beta}\ln (\frac{1 - m - \rho_{s} }{2}) +
   \frac{1}{2\beta}\ln (\frac{1 + m - \rho_{s} }{2})
  \nonumber\\
  & = & 0
 \label{eq:eff_MF_dfdm}\\
    \frac{\partial f}{\partial \rho_{s}} & = &
  \mu
  + X_{0}
  + 4 X_{1} \left( 1 - 2 \rho_{s}  \right)
  + 2 X_{2} {m^2}
  + 6 X_{3} \left( 1 - 3 \rho_{s}  \right)  \left( 1 - \rho_{s}  \right)
  \nonumber\\
  & & \phantom{\mu}
  + 4 X_{6} {m^2} \left( 1 - 2 \rho_{s}  \right)
  + 4 X_{8} {{\left( 1 - \rho_{s}  \right) }^2} \left( 1 - 4 \rho_{s}  \right)
  + X_{9} {m^4}
  \nonumber\\
  & & \phantom{\mu}
  + 2 X_{10} {m^2} \left( 1 - 3 \rho_{s}  \right)  \left( 1 - \rho_{s}  \right)
  + X_{12} {{\left( 1 - \rho_{s}  \right) }^3} \left( 1 - 5 \rho_{s}  \right)
  + 4 K \left( 1 - \rho_{s}  \right)
  \nonumber\\
  & & \phantom{\mu}
  -\frac{1}{2\beta}\ln (\frac{1 - m - \rho_{s} }{2})
  -\frac{1}{2\beta}\ln (\frac{1 + m - \rho_{s} }{2})
  +\frac{1}{\beta}\ln (\rho_{s} )
 \nonumber\\
 & = & 0
 \label{eq:eff_MF_dfdrho}
\eea
The $m$ and $\rho_{s}$ that satisfy these equations
will be the equilibrium values for $m$ and $\rho_{s}$.
There may be several solutions, and the solution that
gives the lowest free energy yields the equilibrium values.

By symmetry the critical line, separating the high-temperature
disordered phase from the low-temperature oil- {\em or}
water-rich phase, is found at $H=m=0$, where equation \ref{eq:eff_MF_dfdm}
is trivially satisfied. Thus, only equation
(\ref{eq:eff_MF_dfdrho}) is important for the localization of the
critical line. At the critical line the minimum changes into
a maximum. The minimal eigenvalue of the matrix of second partial derivatives
of $f$
must therefore vanish. In the present case the matrix has a very simple form:
the nondiagonal elements vanish, and the diagonal element
$\left( \partial^2 f / \partial \rho_{s}^2 \right)$
is
always larger than the element
$\left( \partial^2 f / \partial m^2 \right)$.
Hence, the critical line is defined by the vanishing of the
latter:
\be
  -4 J + \frac{1}{\beta (1 - \rho_{s}) }
  +\left( 4 X_{2} + 8 X_{6} \left( 1 - \rho_{s}  \right)  +
      4 X_{10} {{\left( 1 - \rho_{s}  \right) }^2} \right)  \rho_{s}
  = 0
  \label{eq:eff_MF_critical2}
\ee
This critical condition is independent of the chemical potential,
$\mu$, and it is
therefore, for given interaction strengths and temperature,
an equation for $\rho_{s}$. The equation is a fourth order
polynomal equation, and of the four solutions, only one is
physically acceptable ($\rho_{s}$ real, and
$0<\rho_{s}<1$). This solution is the critical density,
$\rho_{s}^{\mbox{\scriptsize
crit}}(T)$.
The surfactant density and chemical potential are connected
by equation~\ref{eq:eff_MF_dfdrho} giving the critical chemical potential
\bea
  \mu^{\mbox{\scriptsize crit}}(T;\rho_{s})
  & = &
  - 4 K \left( 1 - \rho_{s}  \right)
  -  X_{0}
  - 4 X_{1} \left( 1 - 2 \rho_{s}  \right)
  - 6 X_{3} \left( 1 - 3 \rho_{s}  \right)\left( 1 - \rho_{s}  \right)
  \nonumber\\
  & &
  - 4 X_{8} {{\left( 1 - \rho_{s}  \right) }^2} \left( 1 - 4 \rho_{s}  \right)
  - X_{12} {{\left( 1 - \rho_{s}  \right) }^3} \left( 1 - 5 \rho_{s}  \right)
  \nonumber\\
  & &
  +\frac{1}{\beta}\ln (\frac{1 - \rho_{s} }{2 \rho_{s}})
  \label{eq:eff_MF_critical1}
\eea

The phase transition is not always continuous.
For low temperatures it changes to
first order at a tricritical point, below which the free energy may be lowered
near the ``critical'' line, as defined by equations (\ref{eq:eff_MF_critical2})
and
(\ref{eq:eff_MF_critical1}), by an infinitesimal displacement
$(m,\delta\rho_{s}) = (\epsilon,a\epsilon^2)$, ie a {\em quadratic\/}
relation between $\delta\rho_{s}$ and $m$ ($\delta\rho_{s} \sim m^2$).
All odd partial derivatives with respect to $m$ vanish by symmetry,
and close to the critical line the free energy functional is then:
\bea
  f(m,\delta\rho_{s}) & = & f(0,0)
 +\frac{1}{2}\left(\frac{\partial^2
f}{\partial\rho_{s}^2}\right)\delta\rho_{s}^2
 +\frac{1}{2}
  \left(\frac{\partial^3 f}{\partial
m^2\partial\rho_{s}}\right)m^2\delta\rho_{s}
 +\frac{1}{24}
  \left(\frac{\partial^4 f}{\partial m^4}\right)m^4
 + {\cal{O}}(m^6)
 \nonumber\\
 & = & f(0,0)
 +\left[
  \frac{1}{2}\left(\frac{\partial^2 f}{\partial\rho_{s}^2}\right)a^2
 +\frac{1}{2}
  \left(\frac{\partial^3 f}{\partial m^2\partial\rho_{s}}\right)a
 +\frac{1}{24}
  \left(\frac{\partial^4 f}{\partial m^4}\right)
 \right] m^4
 + {\cal{O}}(m^6)
 \label{eq:eff_MF_fseries}
\eea
The tricritical point is the point where, coming from higher temperatures,
it is first possible to find an $a$ such that the term in
square brackets ($\left[\phantom{A}\right]$)
in this equation (\ref{eq:eff_MF_fseries})
vanishes. This defines the equation
\be
   \left(\frac{\partial^3f}{\partial m^2 \partial\rho_{s}}\right)^2
 - \frac{1}{3}\left(\frac{\partial^4f}{\partial m^4}\right)
              \left(\frac{\partial^2f}{\partial\rho_{s}^2}\right)
 = 0
 \label{eq:eff_MF_tricritical}
\ee
with
\bea
   \left(\frac{\partial^3f}{\partial m^2 \partial\rho_{s}}\right) & = &
   \frac{1}{\beta\left( 1 - \rho_{s}  \right)^{2}}
         + 4 X_{2}
         + 8 X_{6} \left( 1 - 2 \rho_{s}  \right)
         + 4 X_{10} \left( 1 - 3 \rho_{s}  \right) \left( 1 - \rho_{s}  \right)
  \label{eq:eff_MF_d3fdm2drho}\\
   \left(\frac{\partial^4f}{\partial m^4}\right) & = &
  \frac{2}{ \beta\left( 1 - \rho_{s}  \right)^3} + 24 X_{9} \rho_{s}
  \label{eq:eff_MF_d4fdm4}\\
   \left(\frac{\partial^2f}{\partial\rho_{s}^2}\right) & = &
      \frac{1}{\beta\left(1 - \rho_{s}\right)\rho_{s}}
    - 4 K
    - 8 X_{1}
    - 12 X_{3} \left( 2 - 3 \rho_{s}  \right)
    \nonumber\\ & &
    - 24 X_{8} \left( 1 - \rho_{s}  \right)  \left( 1 - 2 \rho_{s}  \right)
    - 4 X_{12} {{\left( 1 - \rho_{s}  \right) }^2}\left( 2 - 5 \rho_{s}
\right)
  \label{eq:eff_MF_d2fdrho2}
\eea
to be satisfied at the tricritical point in addition to
equation~\ref{eq:eff_MF_critical2}. The tricritical density, $\rho_{s\,3}$,
and temperature, $T_3$, as a function of surfactant strength, $A$, and
of the parameter $K$ are given in figure~\ref{fig1.1}. The Lifshitz point
is calculated in section~\ref{sec:correl}. Note that in
earlier mean-field calculations\cite{CHS-4,CHS-7} the
position of the tricritical point was independent of $A$, and
the phase diagram was always of type $C$
for large $A$. In the present calculation this type of phase diagram is found
in an intermediate $A$ range only, and only for $K\approx 0$.
For a sufficiently strong surfactant the phase diagram is
always of type B.


\section{Correlations in high-temperature phase from MF-theory}
  \label{sec:correl}

The connection between the free
energy functional and the spatial correlations is
textbook material\cite{Landau-Lifshitz-1}, and has previously been
applied to mean-field theories of microemulsion
models\cite{CHS-4,CHS-5,CHS-7,Schick-4,Widom-1}:

While the equilibrium values of the fields ($m$ and $\rho_{s}$)
are the same
at all sites, the {\em fluctuations\/} are not
homogeneous.
Expanding to lowest (second) order in the fluctuations,
$\delta m$ and $\delta\rho_{s}$, the free energy takes the form
\bea
  F  =   F_{0}(m,\rho_{s}) +
      \sum_{\Vec{n},\Vec{d}}
       \phi_{\Vec{n}} \Phi(\Vec{d}) \phi^{T}_{\Vec{n}+\Vec{d}}
  \label{eq:MF_corr_F2}
\eea
with
\be
  \Phi(\Vec{d}) = \left( \begin{array}{cc}
            \Phi_{m m}(\Vec{d})        & \Phi_{m \rho_{s}}(\Vec{d})       \\
            \Phi_{\rho_{s} m}(\Vec{d}) & \Phi_{\rho_{s}\rho_{s}}(\Vec{d})
                         \end{array}
                 \right)
  \label{eq:MF_corr_Phi}
\ee
and
\be
  \phi_{\Vec{n}} = \left( \delta m_{\Vec{n}} , \delta\rho_{s\,\Vec{n}} \right)
  \label{eq:MF_corr_phi}
\ee
The Fourier transformed correlation function can then be expressed as
\be
  H(\Vec{k}) = \frac{(2\pi)^2}{\beta} \Phi^{-1}(\Vec{k})
  \label{eq:MF_corr_Hk}
\ee

   In the present short-range model $\Phi(\Vec{k})$ is a
finite sum over near neighbor terms.
We are primarily
interested in the zero field model
($H=0$) where $m=0$. Here $\Phi(\Vec{k})$ is diagonal
(non-diagonal elements are proportional to~$m$) with the elements:
\bea
   \Phi_{m m}(\Vec{k}) & = &
  \BAT{l}
  \frac{1}{\beta(1 - \rho_{s} ) } \CR
    - 2 J \left( \cos (k_{x}) + \cos (k_{y}) \right) \CR
    + \left( 2 X_{2} + 4 X_{6} \left( 1 - \rho_{s}  \right)  +
      2 X_{10} {{\left( 1 - \rho_{s}  \right) }^2} \right)  \rho_{s}
    \left( \cos (2 k_{x}) + \cos (2 k_{y}) \right)
  \EA
\label{eq:MF_corr_Phimm}\\
   \Phi_{\rho_{s} \rho_{s}}(\Vec{k}) & = &
  \BAT{l}
  \frac{1}{\beta\left( 1 - \rho_{s}  \right)  \rho_{s} } \CR
   -\left( \vphantom{X_{2}^{2}} \right.
     \BAT{l}
        2 K + 4 X_{1} + 12 X_{3} \left( 1 - \rho_{s}  \right)  + \CR
      \left.
        12 X_{8} {{\left( 1 - \rho_{s}  \right) }^2} +
        4 X_{12} {{\left( 1 - \rho_{s}  \right) }^3}
      \phantom{x} \right)
        \left( \cos (k_{x}) + \cos (k_{y}) \right)
     \EA \CR
   +\left( 2 X_{3} + 4 X_{8} \left( 1 - \rho_{s}  \right)  +
      2 X_{12} {{\left( 1 - \rho_{s}  \right) }^2} \right)  \rho_{s}
    \left( \cos (2 k_{x}) + \cos (2 k_{y}) \right)  \CR
   +\left( 4 X_{3} + 8 X_{12} \left( 1 - \rho_{s}  \right)  \right)  \rho_{s}
    \left( \cos (k_{x} - k_{y}) + \cos (k_{x} + k_{y}) \right)
   \EA
\label{eq:MF_corr_Phirhorho}
\eea

A detailed discussion of the form of these correlations, and comparison with
experimental and Monte Carlo scattering functions, and other theoretical
work, is deferred to a separate paper.


\subsection{The Lifshitz line}

   The competition between the $\cos(k)$-term
(proportional to the water--water interaction, $J$) and the $\cos(2k)$-term
(proportional to the surfactant density, $\rho_{s}$)
in $\Phi_{m m}(\Vec{k})$ (equation \ref{eq:MF_corr_Phimm}) is the
origin of a maximum in the correlation function,
$H_{m m}(\Vec{k})$, at $\Vec{k} \ne 0$ in parts of
the phase diagram. This maximum is an experimental  characteristic of
microemulsions,
and its location determines the
characteristic length scale for the microemulsion
structure. The maximum is located on
the diagonal ($k_x = k_y= k_{\mathrm{max}}$):
\be
  \cos(k_{\mathrm{max}}) = \frac{J}{\left( 4 X_{2}
                          + 8 X_{6} \left( 1 - \rho_{s}  \right)  +
      4 X_{10} {{\left( 1 - \rho_{s}  \right) }^2} \right)  \rho_{s}  }
  \label{eq:MF_corr_coskmax}
\ee
The Lifshitz line separates the region of the phase diagram where
the correlation function has a maximum at $\Vec{k} \ne 0$
(microemulsion) from
the region with a maximum at $\Vec{k} = 0$
(ordinary fluid). Note that no phase transition
occurs at this line. The Lifshitz line
satisfies the equation
\be
  \cos(k_{\mathrm{max}}) = 1 \; .
  \label{eq:MF_corr_disorder}
\ee


\subsection{Phase transition from disordered to incommensurate}

   Critical behavior is recognized by divergencies in the correlations,
i.e.\ by zeros in $\Phi(\Vec{k})$. The disorder--oil/water
critical line (equation \ref{eq:eff_MF_critical2})
is recovered as $\Phi_{m m}(0) = 0$, while divergencies at
$\Vec{k} \ne 0$ signal a possible continuous transition into
an incommensurate phase (ordered structured phase where
the order is incommensurate with the underlying lattice).
In this section I {\em assume \/} that the transition is
continuous. There is no real justification for this
assumption, and the transition will most probably be
first order in at least part of the phase diagram\cite{CHS-7,Laradji}.

The transition will
take place at $\Vec{k} = (k_{\mathrm{max}},k_{\mathrm{max}})$
(equation \ref{eq:MF_corr_coskmax}), and for a diagonal $\Vec{k}$
the equation $\Phi_{m m}(\Vec{k}) = 0$ is quadratic in $\cos(k)$.
In different regions of the phase diagram the solutions
to this equation fall into one of three classes:

{\bf 1)  Two complex conjugate solutions: \/}No maximum at $\Vec{k}\ne 0$.

{\bf 2) Two real solutions: \/}$\Phi_{m m}(\Vec{k})<0$ for some $\Vec{k}$.
        This is unphysical and the system must be in an ordered phase.

{\bf 3) A single real solution: \/}Divergence at $\Vec{k} \ne 0$,
        i.e.\ critical line.

Using the criterion for class 3 above, the line of continuous
phase transitions into an incommensurate phase satisfies the equation
\be
   b^2 - 8 c (a-c) = 0
  \label{eq:MF_corr_incom}
\ee
with
\bea
  a & = & \frac{1}{\beta(1-\rho_{s})} \nonumber\\
  b & = & 4 J \nonumber\\
  c & = & 4 \rho_{s} \left(
          X_{2} + 2 X_{6}(1-\rho_{s}) + X_{10}(1-\rho_{s})^2
                    \right)  = \frac{J}{\cos(k_{\mathrm{max}})} \nonumber
\eea


\section{Monte Carlo simulation}
\label{sec:MC}

The model has been simulated using a standard Metropolis
Monte Carlo method. The lattice is divided into three
sublattices to enable vectorization,
in a manner very similar to the partition used
by Wansleben et al\cite{Wansleben-1,Wansleben-2}, but no multispin coding is
used.
The random number generator employed is a shift-register
generator with a very long period ($2^{607}-1$), which is
never exhausted in a MC run (For a discussion of
random generators of this type, see the paper of
Compagner and Hoogland\cite{Compagner} and references
therein). The program generates
$\approx 10^6$ spin-flip trials per second on a
CRAY~X-MP. To simplify (and speed up) the program,
I have only simulated the $K=0$, version of the
hamiltonian (\ref{eq:model-hamiltonian}).

Detailed Monte Carlo studies
have been performed for the $K=0, \; A=3J$ model only.

Standard finite size scaling techniques was used to
locate the critical line\cite{Binder}.
This includes
scaling of the magnetization ($m$), susceptibility ($\susc_m$),
and fourth order cumulant, ($U_m$):
\bea
  m & = & \left| \frac{1}{L^2} \sum_i \sigma_i \right|
\label{eq:MC_m}\\
  \susc_m & = & \middel{m^2} - \middel{m}^2
\label{eq:MC_chi_m}\\
  U_m & = & 1 - \frac{\middel{m^4}}{3\middel{m^2}^2}
\label{eq:MC_U_m}
\eea
$L$ is the linear size of the lattice. Since all the critical
exponents are exactly known (the order--disorder
transition is in the universality class of
the two-dimensional Ising model) there is only
one parameter, the critical temperature, $T_c$,
to be fitted in all these scaling
laws:
\bea
  \widetilde{m}(\tau)         & = & L^{1/8} m(\tau)
  \nonumber\\
  \widetilde{\susc}_{m}(\tau) & = & L^{-7/4} \susc_{m}(\tau)
  \nonumber\\
  \widetilde{U}_m(\tau)       & = & U_m(\tau)
\label{eq:MC_crit_scaling}\\
  \tau \XX = \XX \left( \frac{T - T_c}{T_c} \right) L
  \nonumber
\eea
As is evident from figure~\ref{fig2} finite size scaling
works well for all scaling functions, with
a unique critical temperature.

The next question to be answered is whether the phase
transition changes into first order at a tricritical point,
as is the case in the mean-field phase diagrams calculated
above. The Monte Carlo results show pronounced hysteresis
at low temperatures, but this is definitely not conclusive,
and especially not so in this particular model,
since the tricritical temperature is very low.
More convincing is the fact that the
susceptibility, $\susc_s$, of the noncritical surfactant density,
$\rho_s$
\bea
  \rho_s & = & \frac{1}{L^2} \sum_i \left( 1 - \sigma_{i}^2 \right)
\label{eq:MC_rho_s}\\
  \susc_{s} & = & \middel{ \rho_s \rho_s } - \middel{\rho_s}^2
\label{eq:MC_chi_s}
\eea
at the critical line obeys finite size scaling using the
tricritical exponents relevant for the model\cite{Lawrie}
(Universality
class of a one component order parameter in two dimensions,
ie tricritical \spinone Ising model):
\be
  \BA{rcl}
    \widetilde{\susc}_s(\tau) \XX = \XX
      L^{-\gamma_t/\nu_t} \susc_s(\tau) \CR
    \tau \XX = \XX \left( \frac{\mu - \mu_3}{\mu_3} \right) L^{1/\nu_t} \\
    \multicolumn{3}{l}{\mbox{or}} \CR
    \tau \XX = \XX \left( \frac{T - T_3}{T_3} \right) L^{1/\nu_t}
  \EA
  \; .
  \label{eq:MC_tricrit_scale}
\ee
with $\nu_t = 5/9$ and $\gamma_t = 37/36$.
Tricritical finite size
plots are shown in figure~\ref{fig3}. The resulting tricritical point for
$K=0$ and $A=3J$ is $\kb T_3/J = 0.52\pm 0.03$ and $\mu_3/J = 3.15\pm 0.1$.

The Monte Carlo estimates for points along the line of continuous
$\mathrm{oil/water}\to\mathrm{disordered}$ transitions, and of the
tricritical point, for $A=3.0 J$ and $K=0$ are included in
figure~\ref{fig4}. The present calculation is inconsistent with earlier
Monte Carlo calculations by Laradji et al\cite{Laradji}, which gave critical
temperature estimates that are about a factor $1.4$ higher than the
present estimates. I have no explanation for the
discrepancy, but tend to believe that the present estimates are better
than the estimates of Laradji et al.
The reason for this is that in the present calculation the critical temperature
is
always lower than the pure Ising ($\mu=\infty$) value. This is not
the case in the calculations of Laradji et al.
Normally the presence of surfactants will counteract ferromagnetic order,
as it does in the $A=0$ model\cite{Selke}, and it is difficult to see
how it could promote such order.

It was not possible to get conclusive results on the nature of the
$\mathrm{d}\to\mathrm{i}$ (disordered-- to ordered structured phase)
transition.
I will therefore not rule out completely the possibility that a
first-order $\mathrm{d}\to\mathrm{i}$ line crosses the
critical line at a higher temperature than the tricritical point found above.
The model becomes very difficult to study with the present algorithm
close to the tricritical point. In the order of $>10^6$ whole lattice
sweeps were necessary to get reasonable statistics at the points simulated,
and the finite size plots in figure~\ref{fig2} represents
$\approx 20\, \mathrm{hours}$ of Cray CPU-time.
The dynamics grows slower as the tricritical point is
approached. The reason for the
slow dynamics seems to be that the activity is limited to oil/surfactant/water
interfaces while most of the system lies within homogeneous oil
or water regions.


\section{Conclusion}
\label{sec:conc}

Representative phase diagrams are shown in figures~\ref{fig4}
and~\ref{fig5}. Except for weak surfactant, the
phase diagrams are all of type B,
and include an oil/water/microemulsion coexistence region.
This typical behavior was also found in studies of the
Schick model\cite{Schick-4,Schick-5,Schick-7}, and this supports the
claim in section~\ref{sec:spinone} that this simpler
model preserves the important physical properties of
the CHS model. The previous simple mean-field
result\cite{CHS-4,CHS-5,CHS-7} that the phase diagram is always
of type C for strong surfactant, seems
to be an artifact of the approximation employed.

Results from Monte Carlo simulations are consistent with the phase
diagram topologies found from mean-field theory, but the
results are not conclusive. The quantitative error in the
location of the mean-field critical and tricritical temperatures
is of the order expected in two dimensions. Note that the
tricritical temperature found here in the simulations on the
$K=0, \; A=3J$ model, $\kb T_3/J=0.52\pm0.03$ is lower than the
tricritical temperature for the $K=0, \; A=0$ model\cite{Beale},
$\kb T_3/J=0.610\pm0.005$. This confirms
the mean-field result that $T_3$ initially decreases with
larger $A$ (see figure~\ref{fig1.1}).
The critical and tricritical surfactant densities found
in the Monte Carlo simulations are much lower than the
corresponding mean-field densities. The amount of surfactant
necessary to make oil and water miscible is thus much
lower than what mean-field theory predicts.
For the $K=0, \; A=3J$ model the tricritical density
is $\rho_{s\, 3}^{\mathrm{MC}}\approx 0.06$ compared with the mean-field
result $\rho_{s\, 3}^{\mathrm{MF}}=0.2$, and the $A$ independent
$K=0$ tricritical density $\rho_{s\, 3}^{\mathrm{MF}'}=2/3$ of mean-field
theory applied to the CHS hamiltonian directly is totally off
the mark. The Monte Carlo results are compatible with
experimental microemulsion systems where the amount
of surfactant is typically a few percent.
Notice also (figure~\ref{fig4}) that the MC critical line
is almost vertical (constant $\rho_{s}$)
over a wide range of temperatures near the tricritical point.
This feature is not shared by the mean-field critical line
which has an almost constant slope for all temperatures.

The present study strengthen previous results on the same model
that the decisive microscopic ingredient for microemulsion
behavior is the amphiphilic nature of the surfactant.

\acknowledgements

The Norwegian Council for Science and the Humanities (NAVF) is
acknowledged for financial support and CPU-time on the
Cray~X-MP at the supercomputing center in Trondheim.
Several discussions with professor J.~S.~H{\o}ye
have given valuable contributions to the present work.

\unletteredappendix{}

In two dimensions the new effective interactions resulting from the summing out
of the
surfactant directions at a site $i$ involve the factor $(1-\sigma_i^2)$
and all combinations of $\sigma$ and $\sigma^2$ on $i$'s next neighbor sites
that are compatible with the symmetries of the CHS hamiltonian.
Labeling the central site $\sigma_0$ and numbering its nearest neighbor spins
$\sigma_1 \ldots \sigma_4$ in the clockwise direction, these combinations are:
\be
  \BA{l}
    X_0 (1-\sigma_0^2) \CR
    X_1 (1-\sigma_0^2) (\sigma_1^2 + \sigma_2^2 + \sigma_3^2 + \sigma_4^2) \CR
    X_2 (1-\sigma_0^2) (\sigma_1\sigma_3 + \sigma_2\sigma_4) \CR
    X_3 (1-\sigma_0^2) (\sigma_1^2\sigma_3^2 + \sigma_2^2\sigma_4^2) \CR
    X_4 (1-\sigma_0^2) (\sigma_1\sigma_2 + \sigma_2\sigma_3 +
                        \sigma_3\sigma_4 + \sigma_4\sigma_1) \CR
    X_5 (1-\sigma_0^2) (\sigma_1^2\sigma_2^2 + \sigma_2^2\sigma_3^2 +
                        \sigma_3^2\sigma_4^2 + \sigma_4^2\sigma_1^2) \CR
    X_6 (1-\sigma_0^2) (\sigma_1\sigma_3 (\sigma_2^2 + \sigma_4^2) +
                        \sigma_2\sigma_4 (\sigma_1^2 + \sigma_3^2) ) \CR
    X_7 (1-\sigma_0^2) (\sigma_1\sigma_2 (\sigma_3^2 + \sigma_4^2) +
                        \sigma_2\sigma_3 (\sigma_1^2 + \sigma_4^2) +
                        \sigma_3\sigma_4 (\sigma_1^2 + \sigma_2^2) +
                        \sigma_4\sigma_1 (\sigma_2^2 + \sigma_3^2) ) \CR
    X_8 (1-\sigma_0^2) (\sigma_1^2\sigma_3^2 (\sigma_2^2 + \sigma_4^2) +
                        \sigma_2^2\sigma_4^2 (\sigma_1^2 + \sigma_3^2) ) \CR
    X_9 (1-\sigma_0^2) \sigma_1\sigma_2\sigma_3\sigma_4 \CR
    X_{10} (1-\sigma_0^2) (\sigma_1\sigma_2^2\sigma_3\sigma_4^2 +
                        \sigma_1^2\sigma_2\sigma_3^2\sigma_4) \CR
    X_{11} (1-\sigma_0^2) (\sigma_1\sigma_2\sigma_3^2\sigma_4^2 +
                           \sigma_1^2\sigma_2\sigma_3\sigma_4^2 +
                           \sigma_1^2\sigma_2^2\sigma_3\sigma_4 +
                           \sigma_1\sigma_2^2\sigma_3^2\sigma_4) \CR
    X_{12} (1-\sigma_0^2) \sigma_1^2\sigma_2^2\sigma_3^2\sigma_4^2
  \EA
  \label{eq:spinone-2D-terms}
\ee
Considering configurations of four spins around a central surfactant,
the summing out of the surfactant directions leads
to a set of linear equations for the effective couplings, $X_n$.
Solving these equations gives:
\bea
X_{0} &=& -\kbT \ln 4 \nonumber\\
X_{1} &=& -A + 2 \kbT\ln ({2 \over {1 + {e^{-\beta A }}} }) \nonumber\\
X_{2} &=& A - \kbT\ln ({2 \over {1 + {e^{-2 \beta A }}} }) \nonumber\\
X_{3} &=& A - \kbT\ln ({ {8 \left( 1 + {e^{-2 \beta A }} \right) } \over
                        {{{\left( 1 + {e^{-\beta A }} \right)}^4}}
                     }) \nonumber\\
X_{4} &=& 0       \nonumber\\
X_{5} &=& X_{3} \nonumber\\
X_{6} &=& -\half A + \half \kbT\ln ({{4
           \left( 1 + {e^{-2 \beta A }} - {e^{-\beta A }}
           \right) }\over {{{\left( 1 + {e^{-2 \beta A }} \right) }^2}}})
\nonumber\\
X_{7} &=& 0
 \label{eq:spinone-2D-solutions}\\
X_{8} &=& -{3 \over 2}A + \half \kbT\ln ({{1024
           {{\left( 1 + {e^{-2 \beta A }} \right) }^6}}\over
         {\left( 1 + {e^{-2 \beta A }} - {e^{-\beta A }}
         \right)  {{\left( 1 + {e^{-\beta A }} \right) }^{16}}}
         }) \nonumber\\
X_{9} &=& \half A - \quarter \kbT\ln ({ {8 \left( 1 + {e^{-4 \beta A }} \right)
}
                                     \over
                                       {{{\left( 1 + {e^{-2 \beta A }}
                                          \right)}^4}}
                                     }) \nonumber\\
X_{10} &=& \half A - \quarter \kbT\ln ({ {32 {{\left( 1 + {e^{-2 \beta A }} -
                {e^{-\beta A }} \right) }^4}} \over
                                        {\left( 1 + {e^{-4 \beta A }} \right)
                {{\left( 1 + {e^{-2 \beta A }} \right) }^4}}
                                     }) \nonumber\\
X_{11} &=& 0 \nonumber\\
X_{12} &=& {5 \over 2}A - \quarter \kbT\ln ({ {2^{35} \left( 1 + {e^{-4 \beta A
}} \right)
           {{\left( 1 + {e^{-2 \beta A }} \right) }^{28}}}
                                           \over
                                          {{{\left( 1 + {e^{-2 \beta A }} -
                                               {e^{-\beta A }} \right) }^8}
           {{\left( 1 + {e^{-\beta A }} \right) }^{64}}}
                                          }) \nonumber
\eea


\figure{
  Types of phase diagrams that is expected in the
  CHS-model\cite{CHS-4,CHS-7}. $T$ is temperature, and $\mu$ is the
  surfactant chemical potential. type A, {\normalshape (a)},
  is relevant for a weak surfactant:
  two regions, disordered {\normalshape (d)}
  and oil/water coexistence {\normalshape (o-w)} is
  separated by a continuous phase transition (full line) which
  changes into first order (dashed line) at a tricritical point.
  With a strong surfactant, the phase diagram is either of type B,
{\normalshape (b)},
  or type C, {\normalshape (c)}. Both these diagrams show four regions:
disordered
  {\normalshape (d)},
  microemulsion {\normalshape (m)},
  oil/water coexistence {\normalshape (o-w)}, and ordered
  structured phases {\normalshape (i)}, which may be incommensurate with the
underlying
  lattice.
  \label{fig1}
}

\figure{
  Ground states, and $T=0$ phase diagram of CHS-model in two dimensions.
  The model has
  three different ground states: All oil or water {\normalshape
  (F)}, all surfactant {\normalshape (S)}, layered
  {\normalshape (L)}, and tubular {\normalshape (T)}. The
  layered and tubular states, correspond to ordered structured phases
  {\normalshape (i)} in the $T\ne 0$ phase diagrams of
  figures~\ref{fig1}, \ref{fig4}, and~\ref{fig5}.
  \label{fig1.0.1}
}

\figure{
  Tricritical density, $\rho_{s\,3}$, (top) and temperature,
  $T_3$, (bottom) as a function of surfactant strength, $A$, with
  $K=0.0\ldots 1.0$ (Full lines). The dashed lines give
  the position of the Lifshitz points in regions of parameter
  space where the phase diagram is of type C (fig.~\ref{fig1}).
  \label{fig1.1}
}

\figure{
  Finite size scaling plots for the magnetization,
  $\widetilde{m} = L^{1/8} m$, fourth order cumulant,
  $\widetilde{U}_m = U_m$, and susceptibility,
  $\widetilde{\susc}_m = L^{-7/4} \susc_m$ as functions of
  scaled temperature, $\tau = L (T-T_c)/T_c$. At $K=0$,
  $A=3.0 J$ and $\mu = 5.0 J$, with $T_c = 1.265 J$.
  \label{fig2}
}

\figure{
  Tricritical scaling: Finite size scaling plots for the
  surfactant density susceptibility, $\susc_s$, {\em at\/} criticality
  as a function of scaled chemical potential (top) and scaled temperature
  (bottom). At  $K=0$ and $A= 3.0 J$ with $\mu_3 = 3.15$ and $T_3 = 0.52$.
  \label{fig3}
}

\figure{
  Phase diagrams for $K=0$ and three different surfactant strengths.
  The lines are the results of mean-field theory on the effective
  \spinone hamiltonian. Full lines are continuous-- and dashed lines
  first-order transitions. The dotted line is the Lifshitz line.
  Note that the line for the $\mbox{\normalshape d}\to\mbox{\normalshape i}$
  transition is found under the assumption that the phase transition is
   continuous, that the region denoted  by {\normalshape i}
  (incommensurate) is probably divided into several
  different phases separated by first-order phase transitions,
  and that this region is separated from the {\normalshape o-w}
  region by a first-order transition\cite{CHS-4,CHS-7}.
  This is not indicated in the figure.
  For comparison the ($A$ independent) localization of the tricritical
  point from the application of mean-field theory on the CHS-hamiltonian
  directly is indicated (with a $+$). Squares are Monte Carlo results for
  the $\mbox{\normalshape d}\to\mbox{\normalshape o-w}$
  transition ($A=3.0J$ only). Error bars for the critical temperature
  are not shown, but are much smaller than
  the symbols, while the error bars for the critical surfactant
  density are shown. The MC tricritical point is indicated by
  a cross surrounded by an error rectangle ($A=3.0J$ and top row only).
  \label{fig4}
}

\figure{
  Phase diagrams for $K=J$ and three different surfactant strengths.
  The notation follows figure~\ref{fig4}.
  \label{fig5}
}


\begin{references}

\bibitem{CHS-1}
  A.\ Ciach, J.\ S.\ H{\o}ye and G.\ Stell.
  J.\ Phys.\ A {\bf 21} L777 (1988)

\bibitem{CHS-2}
  A.\ Ciach, J.\ S.\ H{\o}ye and G.\ Stell.
  J.\ Chem.\ Phys.\ {\bf 90} 1214 (1989)

\bibitem{CHS-3}
  A.\ Ciach and J.\ S.\ H{\o}ye.
  J.\ Chem.\ Phys.\ {\bf 90} 1222 (1989)

\bibitem{CHS-4}
  A.\ Ciach.
  J.\ Chem.\ Phys.\ {\bf 93} 5322 (1990)

\bibitem{CHS-5}
  A.\ Ciach, J.\ S.\ H{\o}ye and G.\ Stell.
  J.\ Chem.\ Phys.\ {\bf 95} 5300 (1991)

\bibitem{CHS-6}
  L.\ Renlie, J.\ S.\ H{\o}ye, M.\ S.\ Skaf and G.\ Stell.
  J.\ Chem.\ Phys.\ {\bf 95} 5305 (1991)

\bibitem{CHS-7}
  A.\ Ciach.
  J.\ Chem.\ Phys.\ {\bf 96} 1399 (1992)

\bibitem{Laradji}
  M.~Laradji, H.~Guo, M.~Grant, and M.~J.~Zuckermann.
  Phys.~Rev.~A {\bf 44} 8184 (1991)

\bibitem{Hornreich}
  R.~M.~Hornreich, R.~Liebman, H.~G.~Schuster, and W.~Selke.
  Z.~Phys.~B {\bf 35} 91 (1979)

\bibitem{Schick-4}
  G.\ Gompper and M.\ Schick.
  Phys.\ Rev.\ B {\bf 41} 9148 (1990)

\bibitem{Schick-3}
  G.\ Gompper and M.\ Schick.
  Chem.\ Phys.\ Lett.\ {\bf 163} 475 (1989)

\bibitem{Goldstein}
  R.\ E.\ Goldstein.
  J.\ Chem.\ Phys. {\bf 83} 1246 (1985)

\bibitem{Schick-2}
  M.\ Schick and Wei-Heng Shih.
  Phys.\ Rev.\ Lett.\ {\bf 59} 1205 (1987)

\bibitem{Matsen-1}
  M.\ W.\ Matsen and D.\ E.\ Sullivan.
  Phys.\ Rev.\ A {\bf 41} 2021 (1990)

\bibitem{Matsen-2}
  M.\ W.\ Matsen and D.\ E.\ Sullivan.
  Phys.\ Rev.\ A {\bf 44} 3710 (1991)

\bibitem{Schick-1}
  M.\ Schick and Wei-Heng Shih.
  Phys.\ Rev.\ B {\bf 34} 1797 (1986)

\bibitem{Schick-6}
  G.\ Gompper and M.\ Schick.
  Phys.\ Rev.\ Lett.\ {\bf 62} 1647 (1989)

\bibitem{Schick-5}
  G.\ Gompper and M.\ Schick.
  Phys.\ Rev.\ A {\bf 42} 2137 (1990)

\bibitem{Schick-7}
  M.~Schick.
  Physica~A {\bf 172} 200 (1991)

\bibitem{Alexander-1}
  S.\ Alexander.
  J.~Phys.~(Paris) Lett.\ {\bf 39} L-1 (1978)

\bibitem{Halley}
   J.~W.~Halley and A.~J.~Kolan.
   J.~Chem.~Phys {\bf 88}, 3313 (1988)

\bibitem{Landau-Lifshitz-1}
  L.\ D.\ Landau and E.\ M.\ Lifshitz: {\em Statistical Physics \/}
  3rd Edition Part 1. p.\ 350. Pergamon Press 1980.

\bibitem{Widom-1}
  B.\ Widom.
  J.\ Chem.\ Phys.\ {\bf 90} 2437 (1989)

\bibitem{Wansleben-1}
  S.\ Wansleben, J.\ G.\ Zabolitzky and C.\ Kalle.
  J.\ Stat.\ Phys. {\bf 37} 271 (1984)

\bibitem{Wansleben-2}
  S.\ Wansleben.
  Comp.\ Phys.\ Comm.\ {\bf 43} 9 (1987)

\bibitem{Compagner}
  A.\ Compagner and A.\ Hoogland.
  J.\ Comp.\ Phys. {\bf 71} 391 (1987)

\bibitem{Binder}
  K.~Binder and D.~W.~Heermann.
  {\em Monte Carlo simulation in statistical physics.}
  Springer 1988.

\bibitem{Lawrie}
  I.~D.~Lawrie and S.~Sarbach.
  in {\em Phase Transitions and Critical Phenomena\/}
  vol.~9 ed.\ C.~Domb and J.~L.~Lebowitz.
  p.~1. Academic Press 1984.

\bibitem{Selke}
   W.~Selke and J.~Yeomans.
   J.~Phys.~A {\bf 16} 2789 (1983)

\bibitem{Beale}
  P.~D.~Beale
  Phys.~Rev.~B {\bf 33} 1717 (1986)

\end{references}
\end{document}